# PHYSICS-INFORMED NEURAL NETWORKS WITH CURRICULUM TRAINING FOR POROELASTIC FLOW AND DEFORMATION PROCESSES

**Yared W. Bekele**[1]

## ABSTRACT

Physics-Informed Neural Networks (PINNs) have emerged as a highly active research topic across multiple disciplines in science and engineering, including computational geomechanics. PINNs offer a promising approach in different applications where faster, near real-time or real-time numerical prediction is required. Examples of such areas in geomechanics include geotechnical design optimization, digital twins of geo-structures and stability prediction of monitored slopes. But there remain challenges in training of PINNs, especially for problems with high spatial and temporal complexity. In this paper, we study how the training of PINNs can be improved by using an idealized poroelasticity problem as a demonstration example. A curriculum training strategy is employed where the PINN model is trained gradually by dividing the training data into intervals along the temporal dimension. We find that the PINN model with curriculum training takes nearly half the time required for training compared to conventional training over the whole solution domain. For the particular example here, the quality of the predicted solution was found to be good in both training approaches, but it is anticipated that the curriculum training approach has the potential to offer a better prediction capability for more complex problems, a subject for further research.

## INTRODUCTION

Physics-Informed Neural Networks (PINNs) [1] have emerged as a highly active research topic across multiple disciplines in science and engineering. PINNs embed governing equations, often in the form of partial differential equations, directly into the loss function of a neural network. This integration enables PINNs to learn solutions to complex scientific and engineering problems, even where data availability might be sparse. PINNs have been investigated in various disciplines including computational fluid dynamics [2], [3], [4] to model turbulent flows and fluid-structure interaction, material science [5], [6] to predict material properties and optimize design, biomechanics and biomedicine [7], [8] to simulate physiological processes and even in finance and economics [9], [10] for modelling economic systems. PINNs have also been studied in computational geomechanics [11], [12], [13], [14] to model complex physical processes in geological media like soils and rocks.

---

[1] SINTEF AS, Trondheim, Norway, yared.bekele@sintef.no

Despite the promise offered by PINNs for various applications, challenges remain in their training process and generalization capability. Some of these challenges include striking a balance between data fidelity and physical compliance to avoid overfitting, gradient vanishing/exploding during automatic differentiation in case of complex equations, optimization challenges because of the added constraints from physical laws, high computational cost and poor prediction capability for complex systems governed by highly nonlinear/coupled partial differential equations. Various researchers have attempted to address the challenges in the training of PINNs in different ways [15], [16], [17], [18]. Curriculum regularization and curriculum training are promising approaches suggested by the cited researchers to improve the training and prediction capability of PINNs. Curriculum regularization and curriculum training, though sounding similar, are conceptually distinct strategies used to enhance the training of PINNs. *Curriculum regularization* in PINNs refers to a method where the training process gradually introduces complexity in terms of the physics constraints applied to the model. Unlike standard training, which enforces all the physical laws from the start, curriculum regularization starts with simpler, perhaps less stringent, versions of these laws or conditions, and progressively moves towards the full, detailed physical models. This approach helps in managing the training complexity and avoids overwhelming the model with highly complex physical constraints at the initial stages, potentially leading to better convergence and stability in training. *Curriculum training*, on the other hand, involves gradually increasing the difficulty of the training dataset itself, rather than the complexity of the regularization or constraints. The idea is inspired by human learning, where we often start with simple concepts and progressively tackle more complex problems. In the context of neural networks, this might mean starting with easier, perhaps clearer or less noisy data, and progressively introducing more challenging or complex data as the training progresses. The latter i.e. curriculum training is the focus of this paper.

This paper focuses specifically on the application of curriculum-based training to coupled flow and deformation problems in poroelastic media. By doing so, it aims to contribute to the body of knowledge by suggesting a pathway for the development of more accurate and reliable computational models in the field of geomechanical simulations. The paper is structured as follows. We first present the governing equations of poroelasticity as they are commonly introduced in computational geomechanics literature. The PINN model architecture and loss metrics are then defined and presented. An idealized numerical example is then presented wherein the analytical solution and the curriculum training strategy are presented in detail. Results are the presented that compare PINN models trained with and without the curriculum training strategy. Concluding remarks are then presented highlighting the advantages of curriculum training with a mention of points for further research.

# GOVERNING EQUATIONS OF POROELASTICITY

The governing equations of poroelasticity are a combination of the overall mass balance equation, the equilibrium or linear momentum balance equation and the linear elastic constitutive equations for stress-strain relationships.

## *Equilibrium equation*

The equilibrium equation, also called the linear momentum balance equation, for a two-phase (solid matrix and fluid) saturated porous medium under isothermal conditions can in a general form be written as

$$\nabla \cdot \overline{\boldsymbol{\sigma}} + \rho \overline{\boldsymbol{b}} = \boldsymbol{0} \tag{1}$$

where $\overline{\boldsymbol{\sigma}}$ represents the total stress tensor in the porous medium, $\rho$ is the overall density of the porous medium and $\overline{\boldsymbol{b}}$ represents the body forces. For a porous medium, it is beneficial to introduce the concept of effective stress, the component of the stress carried by the soli matrix. The effective stress tensor $\overline{\boldsymbol{\sigma}}'$ as a function of the total stress $\overline{\boldsymbol{\sigma}}$ and the pore fluid pressure $\bar{p}$ is given by

$$\overline{\boldsymbol{\sigma}}' = \overline{\boldsymbol{\sigma}} - \bar{p}\boldsymbol{I} \tag{2}$$

where $\boldsymbol{I}$ is an identity matrix. Using equation (2) in (1) and neglecting body forces results in the simplified equilibrium equation as

$$\nabla \cdot \overline{\boldsymbol{\sigma}}' + \nabla \bar{p} = \boldsymbol{0} \tag{3}$$

For a full description of the equilibrium equation, we need to introduce a constitutive model relating stresses and strains. Assuming isotropic linear elasticity, the effective stress tensor $\overline{\boldsymbol{\sigma}}'$ can be expressed in terms of the infinitesimal strain tensor $\overline{\boldsymbol{\varepsilon}}$

$$\overline{\boldsymbol{\sigma}}' = 2\mu \overline{\boldsymbol{\varepsilon}} + \lambda \operatorname{tr}(\overline{\boldsymbol{\varepsilon}}) \boldsymbol{I} \tag{4}$$

where $\mu$ and $\lambda$ are the Lamé parameters. The infinitesimal strain tensor $\overline{\boldsymbol{\varepsilon}}$ can be expressed in terms of the deformation vector $\overline{\boldsymbol{u}}$ as

$$\overline{\boldsymbol{\varepsilon}} = \frac{1}{2} (\nabla \overline{\boldsymbol{u}} + (\nabla \overline{\boldsymbol{u}})^T) \tag{5}$$

For a two-dimensional problem, the deformation vector is $\overline{\boldsymbol{u}} = \{\bar{u}, \bar{v}\}^T$ where $\bar{u}$ and $\bar{v}$ are the deformations along the x and z directions, respectively. Using this and combining equations (3), (4) and (5) results in the following two equilibrium equations for a two-dimensional case:

$$(\lambda + 2\mu) \frac{\partial^2 \bar{u}}{\partial \bar{x}^2} + \mu \frac{\partial^2 \bar{u}}{\partial \bar{z}^2} + (\lambda + \mu) \frac{\partial^2 \bar{v}}{\partial \bar{x} \partial \bar{z}} + \frac{\partial \bar{p}}{\partial \bar{x}} = 0 \tag{6}$$

*Y.W. Bekele*

$$\mu \frac{\partial^2 \bar{v}}{\partial \bar{x}^2} + (\lambda + 2\mu) \frac{\partial^2 \bar{v}}{\partial \bar{z}^2} + (\lambda + \mu) \frac{\partial^2 \bar{u}}{\partial \bar{x} \partial \bar{z}} + \frac{\partial \bar{p}}{\partial \bar{z}} = 0 \qquad (7)$$

*Mass balance equation*

The mass balance equation for a two-phase (solid matrix and fluid) saturated porous medium under isothermal conditions, obtained from superposition of the mass balance equations for the individual phases, is given by

$$\alpha \nabla \cdot \dot{\bar{\boldsymbol{u}}} + \left( \frac{\alpha - n}{K_s} + \frac{n}{K_f} \right) \frac{\partial \bar{p}}{\partial \bar{t}} + \nabla \cdot \bar{\boldsymbol{w}} = Q \qquad (8)$$

where $\alpha = 1 - K/K_s$ is Biot's coefficient, $\bar{\boldsymbol{u}}$ is the solid deformation vector, $n$ is the porosity, $K_s$ is the bulk modulus of the solid, $K_f$ is the bulk modulus of the fluid, $K$ is the total bulk modulus of the porous medium, $\bar{p}$ is the pore fluid pressure, $\bar{\boldsymbol{w}}$ is the fluid velocity vector and $Q$ is a fluid source or sink term. For a porous medium with incompressible constituents, we have $1/K_s = 1/K_f = 0$ and ignoring any source/sink terms, the mass balance equation reduces to

$$\nabla \cdot \dot{\bar{\boldsymbol{u}}} + \nabla \cdot \bar{\boldsymbol{w}} = 0 \qquad (9)$$

If the fluid flow in the porous medium is assumed to be drivel only by pressure gradients (neglecting the effects of gravity), the fluid velocity can be described by Darcy's law and can be expressed as

$$\bar{\boldsymbol{w}} = -\frac{\boldsymbol{k}}{\gamma_f} \nabla \bar{p} \qquad (10)$$

where $\boldsymbol{k}$ is the hydraulic conductivity matrix and $\gamma_f$ is the unit weight of the fluid. For an isotropic porous medium, the hydraulic conductivity is the same in all directions of flow and the magnitude of the hydraulic conductivity, $k$, will apply to the diagonal components of $\boldsymbol{k}$. Based on equations (9) and (10) and the introduction of the deformation vector $\bar{\boldsymbol{u}}$, the mass balance equation reduces to

$$\frac{\partial}{\partial \bar{t}} \left( \frac{\partial \bar{u}}{\partial \bar{x}} + \frac{\partial \bar{v}}{\partial \bar{z}} \right) - \frac{k}{\gamma_f} \left( \frac{\partial^2 \bar{p}}{\partial \bar{x}^2} + \frac{\partial^2 \bar{p}}{\partial \bar{z}^2} \right) = 0 \qquad (11)$$

*Nondimensionalized poroelastic equations*

The governing equations of poroelasticity in two dimensions are those given in equations (6), (7) and (11). The field variables are the deformations $\bar{u}$ and $\bar{v}$ and the pore pressure $\bar{p}$. For implementation into a PINN, it is more convenient to nondimensionalize the governing equations. Introducing a reference length $l$ for the porous medium, we define the nondimensionalized spatial and deformation variables as

$$x = \frac{\bar{x}}{l}, \quad y = \frac{\bar{y}}{l}, \quad u = \frac{\bar{u}}{l} \quad \text{and} \quad v = \frac{\bar{v}}{l} \tag{12}$$

The corresponding nondimensionalized time and pore pressure variables can be derived and shown to be given by

$$t = \frac{(\lambda + 2\mu)k}{\gamma_f l^2} \cdot \bar{t} \quad \text{and} \quad p = \frac{\bar{p}}{\lambda + 2\mu} \tag{13}$$

Using equations (12) and (13) in the governing equations (6), (7) and (11), we get the nondimensionalized poroelastic equations as

$$f(x,z,t) = (\eta+1)\frac{\partial^2 u}{\partial x^2} + \frac{\partial^2 u}{\partial z^2} + \eta\frac{\partial^2 v}{\partial x \partial z} + (\eta+1)\frac{\partial p}{\partial x} = 0 \tag{14}$$

$$g(x,z,t) = \frac{\partial^2 v}{\partial x^2} + (\eta+1)\frac{\partial^2 v}{\partial z^2} + \eta\frac{\partial^2 u}{\partial x \partial z} + (\eta+1)\frac{\partial p}{\partial z} = 0 \tag{15}$$

$$h(x,z,t) = \frac{\partial^2 u}{\partial t \partial x} + \frac{\partial^2 v}{\partial t \partial z} - \frac{\partial^2 p}{\partial x^2} - \frac{\partial^2 p}{\partial z^2} = 0 \tag{16}$$

where the nondimensional parameter $\eta$ is given by

$$\eta = 1 + \frac{\lambda}{\mu} \tag{17}$$

# PHYSICS-INFORMED NEURAL NETWORK MODEL

## *Model architecture*

The model architecture used for the PINN model is a fully connected network where the number of layers and hidden units per layer are optimized during training. For the two-dimensional governing equations derived earlier, the PINN model accepts the spatial and temporal parameters $\{x, z, t\}$ as inputs and predicts the field variables $\{\hat{u}, \hat{v}, \hat{p}\}$ as outputs. Figure 1 shows an illustration of the model architecture. An important component of the PINN model is evaluation of the PDEs based on the predicted values of the field variables. This is performed using *automatic differentiation* which keeps track of the input variables and associated weights throughout the model. TensorFlow is used for implementation of the model and automatic differentiation is performed by defining custom training steps and gradient tapes.

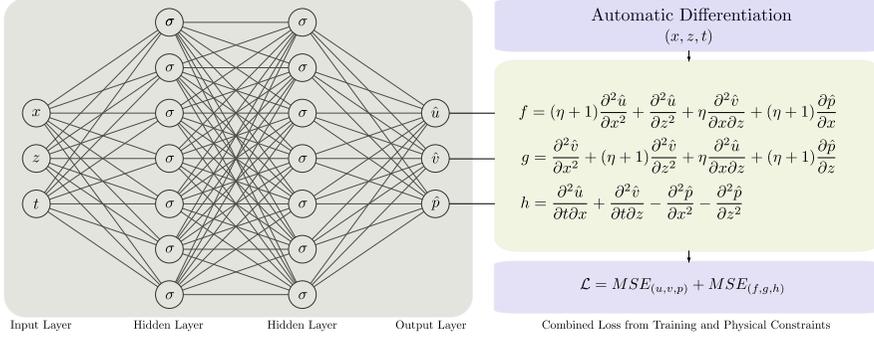

*Figure 1 PINN model architecture. Note that the number of layers and units per layer is only for illustration purposes.*

## Loss metrics

The performance of the PINN model is measured by evaluating the losses both from the data-driven part of the training and the physics loss resulting from applying the PDEs as constraints to the model. The training data involves initial and boundary conditions for the specific problem under consideration. Mean squared error (MSE) is chosen as the metrics for calculating the data loss for each predicted field variable. Based on the actual field variable training data $\{u, v, p\}$ and their corresponding predicted values $\{\hat{u}, \hat{v}, \hat{p}\}$, the losses for each variable are calculated from

$$MSE_u = \frac{1}{N}\sum_{i=1}^{N}(u_i - \hat{u}_i)^2, \ MSE_v = \frac{1}{N}\sum_{i=1}^{N}(v_i - \hat{v}_i)^2 \ \text{and} \ MSE_p = \frac{1}{N}\sum_{i=1}^{N}(p_i - \hat{p}_i)^2 \quad (18)$$

where $N$ represents the number of training data points. The total data loss is calculated as the sum of the individual data losses for each variable i.e.

$$MSE_{data} = MSE_u + MSE_v + MSE_p \quad (19)$$

The performance of the PINN model with respect to obeying the governing PDEs is measured by the mean squared error loss for each PDE. The physics losses are given by

$$MSE_f = \frac{1}{N_c}\sum_{i=1}^{N_c} f_i^2, \quad MSE_g = \frac{1}{N_c}\sum_{i=1}^{N_c} g_i^2 \quad \text{and} \quad MSE_h = \frac{1}{N_c}\sum_{i=1}^{N_c} h_i^2 \quad (20)$$

where $N_c$ is the number of collocation points $(x_c, z_c, t_c)$ where the PDEs are evaluated at. Note that residual terms, if there are any, need to be subtracted from each function during the evaluation of these losses. The total physics loss is the sum of the individual losses from each governing equation i.e.

$$MSE_{physics} = MSE_f + MSE_g + MSE_h \quad (21)$$

The total loss for the PINN model is the sum of the data and physics losses:

$$MSE_{total} = MSE_{data} + MSE_{physics} \quad (22)$$

The PINN model is trained to minimize this loss by adjusting the hyperparameters of the model, such as number of layers, number of units per layer, learning rate and batch size.

## NUMERICAL EXAMPLE

### Analytical solution

To demonstrate the curriculum training strategy, we first generate manufactured analytical solutions for the field variables $u$, $v$ and $p$ that resemble a realistic two-dimensional geotechnical problem. The generated analytical solutions are:

$$u(x,z,t) = x \cdot (1 - e^{-\alpha \cdot z}) \cdot t \cdot e^{-\delta \cdot t} \quad (23)$$

$$v(x,z,t) = (1 - e^{-\beta \cdot z}) \cdot t^2 \cdot e^{-\epsilon \cdot t} \quad (24)$$

$$p(x,z,t) = 3z \cdot (1-z) \cdot e^{-\zeta \cdot t} \quad (25)$$

where $\alpha, \beta, \delta, \epsilon$ and $\zeta$ are parameters that can be adjusted to generate various solutions. These manufactured analytical solutions are used to generate solutions for the field variables on a square domain with nondimensional $x$ and $z$ values ranging from 0 to 1 and nondimensional time $t$ also ranging from 0 to 1. The parameters used to generate the solutions are $\alpha = 0.5, \beta = 2.0, \delta = 1.0, \epsilon = 1.0, \zeta = 1.5$ and $\eta = 2.5$. Figure 2 shows the analytical solution at $t = 0.5$ while Figure 3 shows a profile plot at $x = 0.5$ at selected time steps for the field variables $u$, $v$ and $p$.

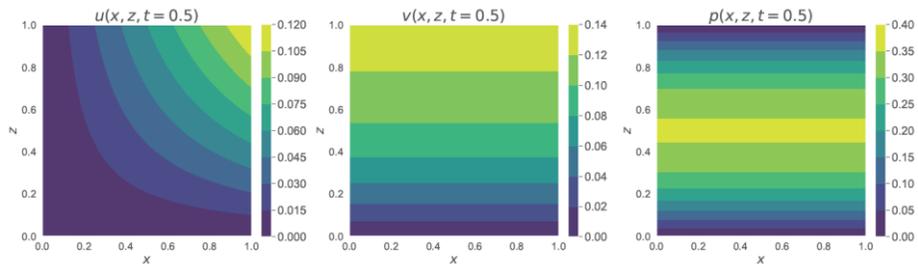

*Figure 2 Analytical solution for u, v and p on the spatial domain at t=0.5.*

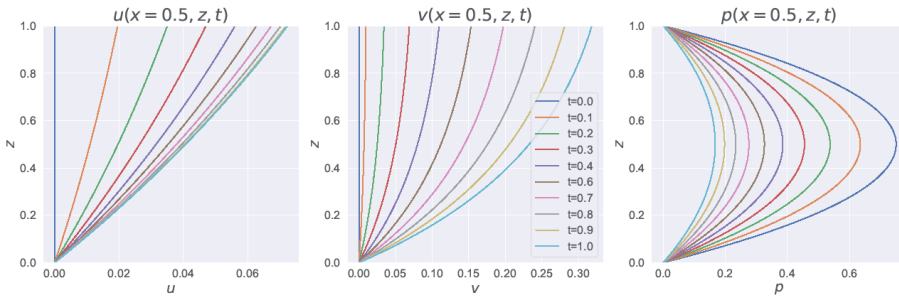

*Figure 3 Profile plots of the analytical solution at x=0.5 for u, v and p.*

These analytical solutions will not directly satisfy the governing PDEs and thus will result in residuals that need to be accounted for. The residuals can be derived by substituting the analytical solutions in Equations (23)-(25) into the governing PDEs in Equations (14)-(16). The resulting residuals corresponding to each PDE are presented in the Appendix.

## *Curriculum training strategy*

Curriculum training is a learning strategy inspired by the idea of learning progressively, analogous to a human student learning a subject by gradually tackling more complex concepts. This approach, also known as curriculum learning, starts with simpler tasks or a subset of the data and progressively increases the difficulty level, either by including more complex examples or by expanding the data range. This methodology has shown effectiveness in various domains, such as natural language processing and computer vision, often leading to faster convergence and better generalization.

In the context of PINNs, curriculum training can be particularly advantageous when dealing with problems with complex physical processes. PINNs leverage the underlying physics of the problem, encoded in the neural network through the differential equations, to predict the field variables. For the specific case of time-dependent problems, like the poroelasticity problem we are dealing with here, curriculum training can be implemented by dividing the time range into intervals and training the model sequentially on these intervals. This strategy aligns with the temporal evolution of physical processes, allowing the model to first learn the initial behavior before advancing to more complex states as time progresses.

Figure 4 highlights the distinction between a standard training approach and one utilizing curriculum training. The plots show how initial and boundary condition data are extracted for training the PINN model using the standard approach and using curriculum training. The left plot shows the conventional approach where the model is trained on the initial and boundary condition data for the entire time domain. This can be challenging, as the model must learn to predict the complex dynamics across the entire time range at once. The right plot, however, demonstrates the curriculum training approach. Here, the time domain is discretized into 10 intervals, where training data for the intervals is extracted separately. Training begins with the first interval, closest to the initial conditions. Gradually, subsequent intervals are introduced, scaffolding the learning process as the complexity of the solution increases with time.

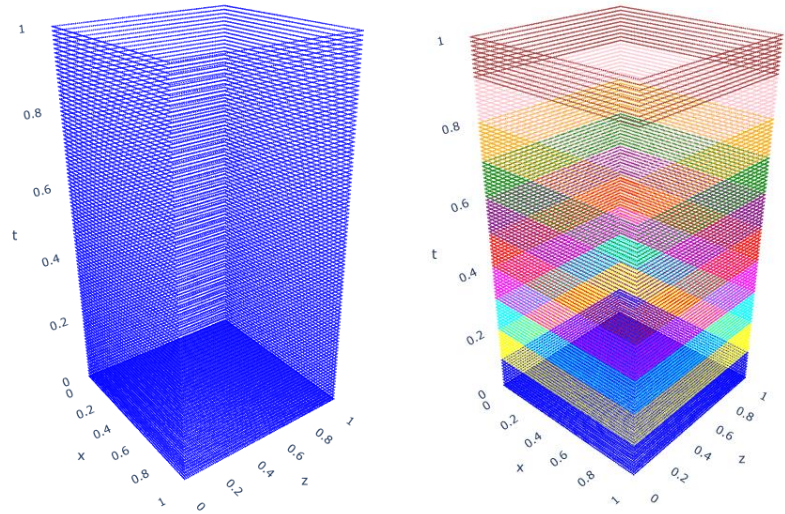

*Figure 4 Extraction of initial and boundary condition training data for the poroelasticity problem. The plot on the left shows extraction of IC and BC data without curriculum training and the plot on the right shows extraction with curriculum training where the time range is divided into 10 intervals.*

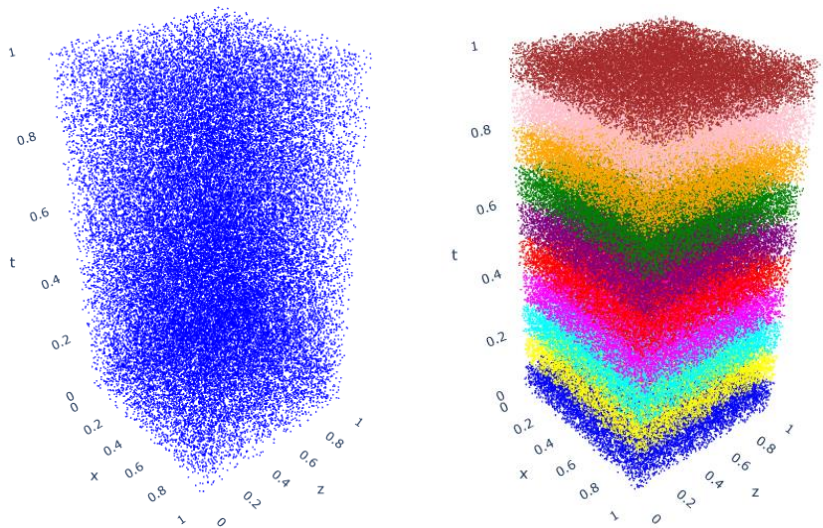

*Figure 5 Generation of collocation points to evaluate the governing PDEs. The plot on the left shows generation of collocation data for the entire solution domain according to the standard training approach while the plot on the right shows generation of collocation points for the different training intervals for curriculum training.*

Figure 5 compares two different methods for generating collocation points, which are used for evaluating the governing PDEs during the training of PINNs using automatic differentiation. The left plot on the figure shows collocation points scattered throughout the entire solution domain, generated using Latin hypercube sampling, for the standard training approach. This requires the model to tackle the complexity of the PDEs on the entire domain from the start, which can be quite demanding. On the right, the collocation points are generated progressively throughout the solution domain, matching the segmented intervals used for data extraction in curriculum training. This approach allows the model to focus evaluating the PDEs and optimizing the model gradually at different solution intervals, usually starting from simple to more complex sections. This helps to streamline the learning process, potentially leading to a more robust model as we will demonstrate in the following sections.

### Training data with and without curriculum

The analytical solution is generated over the domain for a spatial discretization of $N_x \times N_z = 50 \times 50$ and a temporal discretization of $N_t = 50$. This implies a total dataset size of 125 000 for each of the input and output variables, $(x, z, t)$ and $(u, v, p)$, respectively. But only the initial and boundary condition data are used for data-driven training while the PDEs are evaluated at randomly generated collocation points. After extraction of the initial conditions (2500 data points) and the boundary conditions (9800 data points), the total size of the data for training becomes 12 300 for each of the input and output variables.

**Training without curriculum**: For model training without the curriculum training strategy, the initial and boundary condition data throughout the domain are used at once. The batch size is adjusted for optimal training and the data is shuffled during training. The Latin Hypercube Sampling (LHS) method is used to generate collocation points in the solution domain. The number of collocation points generated in this case is 1000.

**Training with curriculum**: For model training with the curriculum training strategy, the temporal domain is divided into the desired number of intervals. For our example here, it is divided into 10 equally spaced intervals. For example, the first interval involves initial and boundary condition data of size 250 and 980, respectively. For subsequent intervals, only boundary condition data of size 980 are relevant. The number of collocation points generated for each interval is 100, implying a total number of collocation points of 1000 as in the case of training without curriculum.

### Model hyperparameters

After some fine tuning, the following model hyperparameters are selected for the PINN model. The number of hidden layers and hidden units per layer is chosen as 5 and 20, respectively, while the *tanh* activation function is chosen. A batch size of 256 is selected to be optimal. Adam optimizer is used with a learning rate of 0.001. These hyperparameters apply to model training both with and without the curriculum training strategy.

### Results

Model training is performed for 3000 epochs for both model training with and without the curriculum training strategy. The loss versus epoch plot for model training without curriculum is shown in Figure 6. The corresponding plots for model training with curriculum are shown in Figure 7. For the particular numerical example here, the loss decreases optimally with the number of epochs in both cases, but the model takes a longer time to train in the first case i.e. model training without the curriculum training strategy. For the selected model hyperparameters and number of epochs, model training with the curriculum training approach takes approximately 3 hours on a laptop with a Core i7 CPU while a similar model without curriculum training takes nearly 6 hours on the same hardware.

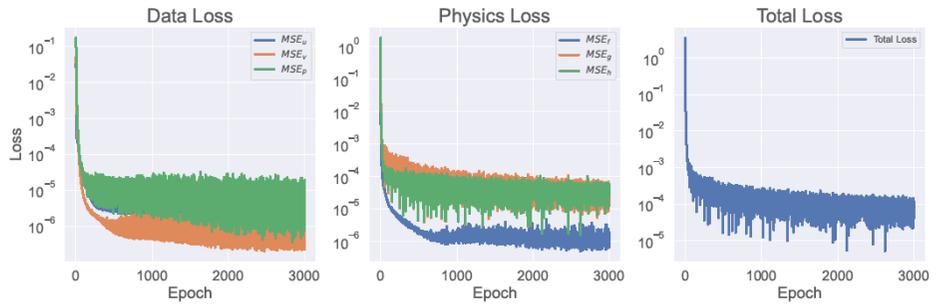

*Figure 6 Data loss, physics loss and total loss versus epoch plots for model training without the curriculum training strategy.*

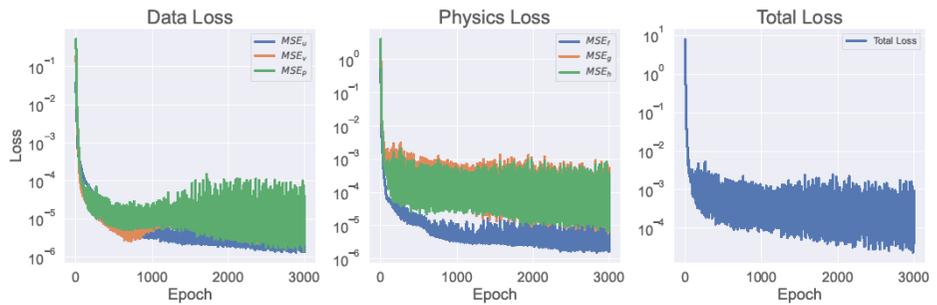

*Figure 7 Data loss, physics loss and total loss versus epoch plots for model training with the curriculum training strategy.*

Figure 8 shows a comparison between the analytical versus predicted solutions over the solution domain for model training without the curriculum training strategy. The results are shown for the solution of the field variables $u, v$ and $p$ at $t = 1.0$. The corresponding results for model training with the curriculum training approach are shown in Figure 9. The predicted solutions in both cases were found to be in good agreement with the analytical solutions.

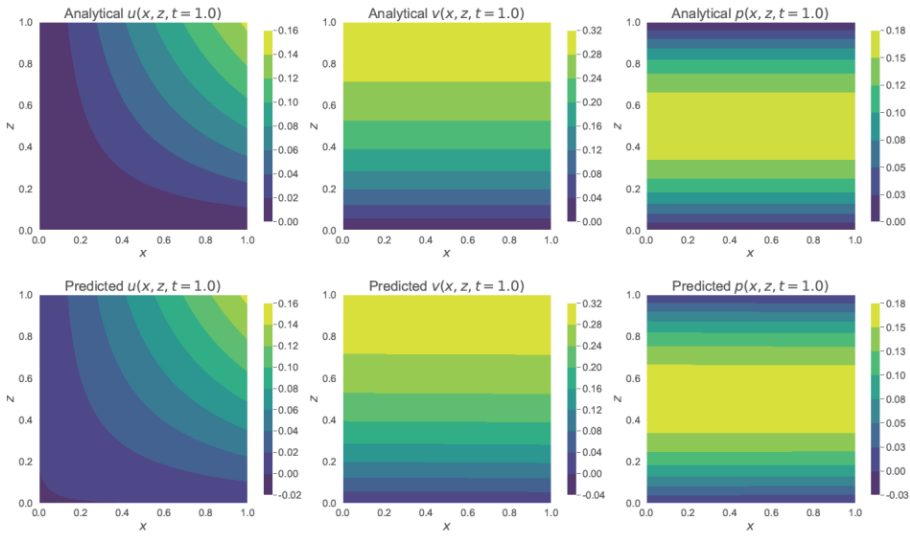

*Figure 8 Analytical versus predicted solutions for the field variables u, v and p at t=1.0 for model training without curriculum.*

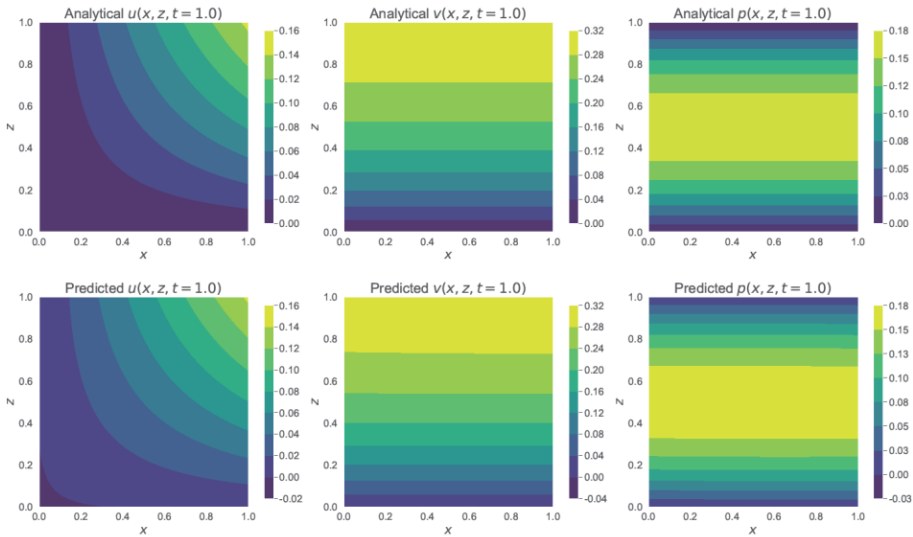

*Figure 9 Analytical versus predicted solutions for the field variables u, v and p at t=1.0 for model training with curriculum.*

To examine the quality of the predicted solutions, we look at the profiles of the field variables along the $z$ direction for a selected $x$ value and selected time steps. The resulting profiles are shown in Figure 10 and Figure 11 for model training without and with curriculum, respectively. In both figures, the profiles are plotted at x=1.0 and for a time range between 0 and 1 with an interval of 0.1.

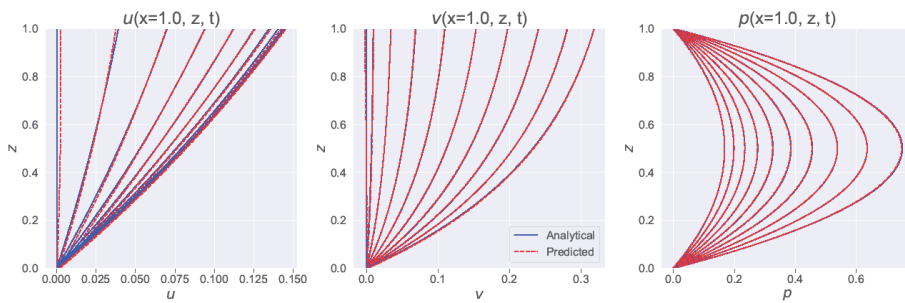

*Figure 10 Analytical versus predicted solution profiles for the field variables u, v and p at x=1.0 for model training without curriculum. The profiles are shown at t intervals of 0.1 from 0 to 1.*

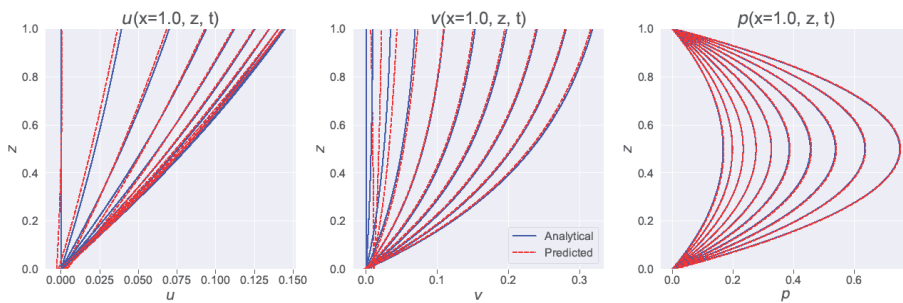

*Figure 11 Analytical versus predicted solution profiles for the field variables u, v and p at x=1.0 for model training with curriculum. The profiles are shown at t intervals of 0.1 from 0 to 1.*

In the numerical example discussed, both training approaches—using a curriculum training strategy and without it—yield reasonably accurate predictions. However, the primary distinction lies in the training duration. Training with the curriculum strategy is observed to be nearly twice as fast as the conventional method under the same model setup and using identical hardware. This efficiency presents a considerable advantage, especially when training PINNs for complex problems. In scenarios involving significant temporal variations within the solution domain, the curriculum-based approach is likely favored. It is anticipated that this method not only enhances training efficiency but also improves the accuracy of the predictions. Further exploration and demonstration of this hypothesis on complex temporal problems will be pursued in future work.

## CONCLUDING REMARKS

The research trend on PINNs in various science and engineering disciplines is briefly highlighted. While PINNs offer great promise for various practical applications, there remain several challenges in training PINN models and reaping their potential practical benefits for faster, near real-time or real-time numerical prediction. Of the various challenges that need to be addressed regarding the training of PINNs, we focused on reducing the high computational cost of training PINN models and improving the quality of the predicted solutions they offer. For this purpose, a curriculum training strategy is employed for developing a PINN model for coupled poroelastic flow and deformation processes. An idealized numerical example is used for to demonstrate the curriculum training strategy where the PINN model is trained gradually by introducing the training data at intervals along the temporal dimension of the solution domain. In line with the different temporal intervals, the governing PDEs are evaluated at collocation points corresponding only to that interval. The curriculum training approach is compared with a conventional training approach where the training data over the whole spatial and temporal domain of the solution is used at once for training and the PDEs are evaluated at collocation points over the whole solution domain. The results show that the curriculum training approach enables training of the PINN model nearly two times faster than the conventional approach for a similar model setup and on the same hardware. For the particular example here, the predicted solutions for both training approaches were found to be in good agreement with the reference solution. It is however anticipated that the curriculum training approach, in addition to enabling faster model training, could result in a superior prediction capability for problems with more temporal complexity in the solution. This will be a subject for further research.

## APPENDIX

The residuals corresponding to the analytical solutions and the respective PDEs are:

$$r_u(x,z,t) = -\alpha^2 \cdot t \cdot x \cdot e^{-\alpha \cdot z - \delta \cdot t}$$

$$r_v(x,z,t) = \left(\alpha \cdot \eta \cdot t \cdot e^{\beta \cdot z + \epsilon \cdot t + t \cdot \zeta} - \beta^2 \cdot t^2 \cdot (\eta + 1) \cdot e^{\alpha \cdot z + \delta \cdot t + t \cdot \zeta} \right.$$
$$\left. - 3 \cdot (\eta + 1) \cdot (2z - 1) \cdot e^{\alpha \cdot z + \beta \cdot z + \delta \cdot t + \epsilon \cdot t}\right)$$
$$\cdot e^{-\alpha \cdot z - \beta \cdot z - \delta \cdot t - \epsilon \cdot t - t \cdot \zeta}$$

$$r_p(x,z,t) = -\beta \cdot t \cdot (\epsilon \cdot t - 2) \cdot e^{-\beta \cdot z} e^{-\epsilon \cdot t} - (1 - e^{-\alpha \cdot z})$$
$$\cdot (\delta \cdot t - 1) \cdot e^{-\delta \cdot t} + 6 \cdot e^{-\zeta \cdot t}$$